# Evidence for a Fermi liquid in the pseudogap phase of high-$T_c$ cuprates

S. I. Mirzaei[1*], D. Stricker[1*], J. N. Hancock[1], C. Berthod[1], A. Georges[1,2,3], E. van Heumen[1,4], M. K. Chan[5], X. Zhao[5], Y. Li[6], M. Greven[5], N. Barišić[5,7,8], and D. van der Marel[1]

[1]Département de Physique de la Matière Condensée, Université de Genève, 24 quai Ernest-Ansermet, CH-1211 Genève, Switzerland.
[2]Centre de Physique Théorique, École Polytechnique, CNRS, 91128 Palaiseau, France.
[3]Collège de France, 11 place Marcelin Berthelot, 75005 Paris, France.
[4]Van der Waals-Zeeman Instituut, Universiteit van Amsterdam - NL-1018XE Amsterdam, the Netherlands.
[5]School of Physics and Astronomy, University of Minnesota, Minneapolis, Minnesota 55455, USA.
[6]International Center for Quantum Materials, School of Physics, Peking University, Beijing 100871.
[7]Institute of Physics, Bijenička c. 46, HR–10000 Zagreb, Croatia.
[8]Service de Physique de l'Etat Condensé, CEA-DSM-IRAMIS, F 91198 Gif-sur-Yvette, France.

Cuprate high-$T_c$ superconductors on the Mott-insulating side of "optimal doping" (with respect to the highest $T_c$'s) exhibit enigmatic behavior in the non-superconducting state. Near optimal doping the transport and spectroscopic properties are unlike those of a Landau-Fermi liquid[1,2]. For carrier concentrations below optimal doping a pseudogap removes quasi-particle spectral weight from parts of the Fermi surface[3], and causes a break-up of the Fermi surface into disconnected nodal and anti-nodal sectors. Here we show that the near-nodal excitations of underdoped cuprates obey Fermi liquid behavior. Our optical measurements reveal that the dynamical relaxation rate $1/\tau(\omega,T)$ collapses on a universal function proportional to $(\hbar\omega)^2+(1.5\pi k_B T)^2$. Hints at possible Fermi liquid behavior came from the recent discovery of quantum oscillations at low temperature and high magnetic field in underdoped YBa$_2$Cu$_3$O$_{6+d}$[4] and YBa$_2$Cu$_4$O$_8$[5], from the observed $T^2$-dependence of the DC ($\omega=0$) resistivity for both overdoped and underdoped cuprates[6,7], and from the two-fluid analysis[8] of nuclear magnetic resonance data[9]. However, the direct spectroscopic determination of the energy dependence of the life-time of the excitations -provided by our measurements- has been elusive up to now. This observation defies the standard lore of non-Fermi liquid physics in high $T_c$ cuprates on the underdoped side of the phase diagram.

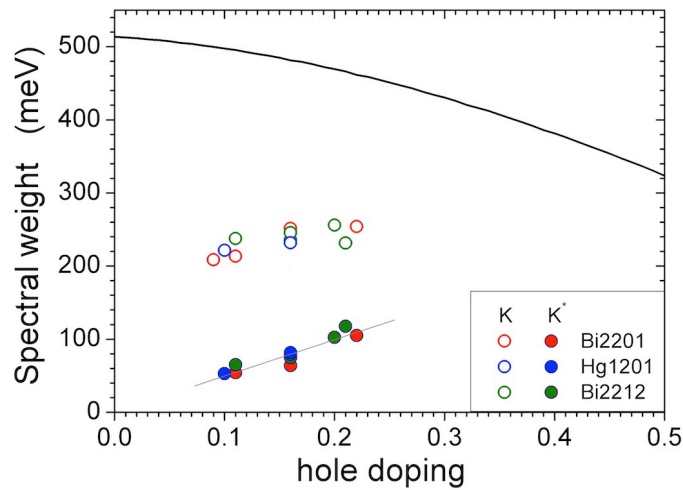

Figure 1 | Conduction band spectral weight per CuO$_2$ layer for a large number of cuprate superconductors. Closed symbols: coherent spectral weight, K*. The grey line is a linear least squares fit, $K^*=496\,x$, where $x$ is the nominal hole-doping. Open symbols: total spectral weight, K. New data are presented for for Hg1201 with 10% doping (see Supplementary Information table S1). For the other materials see table 1 in Ref. 11. The black solid curve represents $K$ calculated from *ab initio* band parameters for Hg1201 (see Appendix VI). For the Bi$_2$Sr$_2$CuO$_6$ (Bi2201) sample with hole doping below 0.10 $K^*$ could not be calculated, since in this case we obtain $M_1(\omega,T)<0$ for $\omega \to 0$, possibly due to intrinsic or stoichiometry-related inhomogeneity.

---

We will analyze in this paper two properties of the interacting electron-system, each of which can be obtained in a straightforward manner from the experimentally measured optical conductivity. These are the dynamical relaxation rate $1/\tau(\omega,T)$, which is equal to the imaginary part of the memory function $M(\omega,T)=M_1(\omega,T)+iM_2(\omega,T)$ [10] and the optical spectral weight of the conduction band electrons, $K$. The expression relating between these two quantities to the optical conductivity of the $CuO_2$ layers is (see Appendix VII)

$$G(\omega,T) = \frac{i\pi K}{\hbar\omega + M(\omega,T)} G_0 \qquad (1)$$

where $G_0=2e^2/h$ is the conductance quantum. The spectral weight $K$ corresponds to minus the kinetic energy if the frequency integration of the experimental data is restricted to intraband transitions. The effect of electron-electron interactions and coupling to collective modes is described by the memory function $M(\omega,T)$. The doping dependence of $K$ and the coherent spectral weight, defined as $K^*= K/(1+M_1(\omega,T)/\hbar\omega)|_{\omega=0}$, is summarized in Fig. 1 for a number of hole-doped cuprates. The theoretical values of $K$ based on the band-parameters obtained from LDA *ab-initio* calculations are about a factor two larger than the measured values, which is due to strong correlation predicted by the Hubbard model for $U/t \geq 4$ [12]. $K$ decreases when the hole doping decreases, but does not extrapolate to zero for zero doping in accordance with the analysis of Comanac *et al.* [13]. In contrast, the coherent spectral weight, $K^*$, is proportional to the hole doping $x$: $K^*=xK^0$, where $K^0=496$ meV, in agreement with the trend observed for $La_{2-x}Sr_2CuO_4$ [14] and $YBa_2Cu_3O_{7-d}$ [15]. This provides strong evidence that a Mott insulator is approached as the doping is reduced. Whether this happens by (i) the quasi-particle residue being gradually suppressed[16,17] or (ii) the Fermi surface arcs shrinking to zero without vanishing of the nodal spectral weight[3,7], are two open possibilities which cannot be decided from these data.

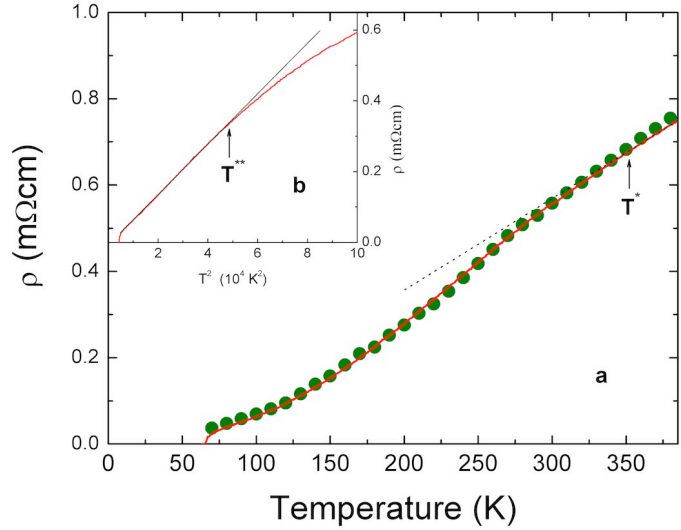

Figure 2 | DC resistivity of underdoped Hg1201.
Inverse optical conductivity extrapolated to zero-frequency (green circles) compared with measured DC-resistivity[7] (solid red line) of a sample of the same composition and doping (a). The dashed black curve is a $T$-linear fit to the resistivity data above 350K. b: The same data as a function of $T^2$, shown with a $T^2$ fit to the data below 220 K. The definitions and values for T* and T** of Ref. 7 are used.

The compound $HgBa_2CuO_{4+\delta}$ (Hg1201) is the single-layer cuprate exhibiting the highest $T_c$'s (97 K). We measured the optical conductivity of strongly under-doped single crystals of Hg1201 ($T_c$=67 K). In the $\omega\rightarrow 0$ limit (see Appendix II) the optical conductivity corroborates the recently reported temperature dependences of the DC resistivity[7] as shown in Fig. 2. Since $K$ is practically temperature independent in the normal state[18], the low temperature $T^2$ dependence of the resistivity is due to the quadratic temperature variation of $M_2(0,T)$. The infrared data confirm that Hg1201 exhibit the lowest residual resistance among the cuprates and a change to a linear temperature dependence above $T^*$ associated with the sudden opening of a pseudogap[19,20]. In the inset of Fig. 2 this is seen as a clear departure from the $T^2$ curve at approximately $5\cdot 10^4$ K$^2$. The DC transport data, owing to the higher precision, allow for Hg1201 crystals of the same composition and doping to identify $T^*\sim 350$K as the temperature above which the resistivity has a linear temperature dependence, and $T^{**}\sim 220$K as the temperature below which the temperature dependence is purely quadratic. Finally superconducting fluctuations become noticeable at $T'\sim 85$K.

The real and imaginary parts of the memory function of underdoped Hg1201 with $T_c$=67 K are shown in Fig. 3 for temperatures from 10 to 390 K. $M_1(\omega,T)$ has a linear slope extrapolating to $\omega$=0, which becomes less steep at higher temperatures. The maximum at 105 meV erodes gradually as temperature increases, but a residue of this structure remains visible even at 390 K. In the lower left inset we show a plot of the frequency and temperature dependent mass enhancement factor, $m^*(\omega,T)=M_1(\omega,T)/\hbar\omega +1$. For $\omega$ above 60 meV $m^*(\omega,T)$ is a monotonously decreasing function of temperature. For $\omega$ not too large, $m^*(T)$ is roughly speaking a plateau at low temperatures, terminating in a weak maximum at $T(max)$ and followed by a linear-like decrease at higher temperature. $T(max)$ increases when $\omega$ decreases and for $\omega\rightarrow 0$ extrapolates to 212 K~$T^{**}$, indicating another way of identifying $T^{**}$. The increase of $m^*(\omega,T)$ from about 3 at 390 K to 5 at $T^{**}$, taken together with the strong temperature dependence of $M_1(\omega,T)$ near its maximum at 105 meV, indicate that the charge carriers become increasingly renormalized when the temperature decreases. Our results also corroborate the observation in Ref. 21, that the integrated optical conductivity does not decrease when $T$ decreases, so that no opening of an optical pseudogap is seen when, at $T^*$, part of the Fermi-surface is removed by a pseudo-gap, despite the emergence at this temperature of a novel ordered state with two Ising-like magnetic collective modes at 54 and 39 meV as observed with inelastic neutron scattering[22].

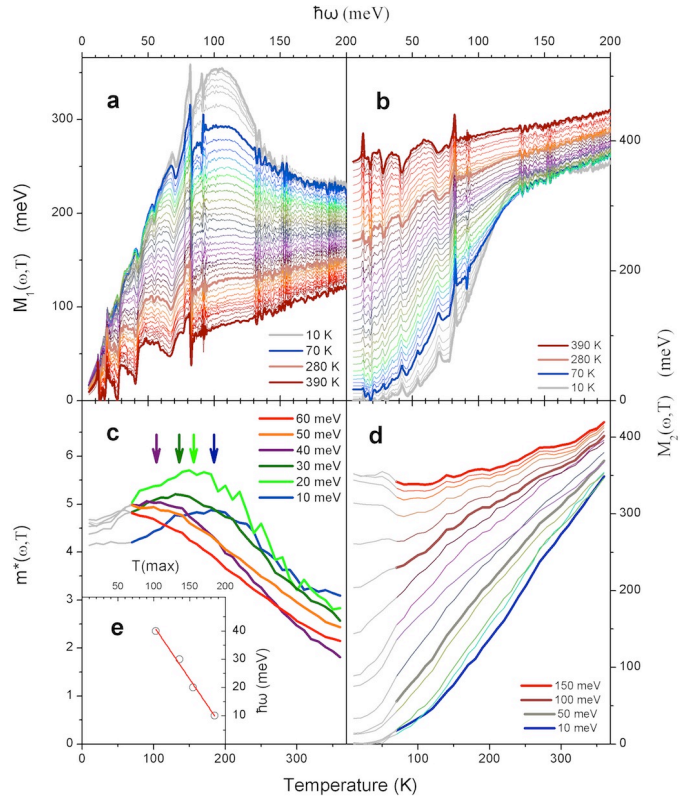

Figure 3 | Optical self energy as a function of frequency and temperature.
Real (a) and imaginary part (b) of the memory function as a function of $\hbar\omega$ for underdoped Hg1201 ($T_c$=67 K). Spectra are shown in 10 K intervals for temperatures from 10 to 390 K. Thick lines are used to highlight the 10 K, 70 K, 280 K and 390 K data. Panel c (d) Effective mass $m^*(\omega,T)$ (relaxation rate $M_2(\omega,T)$) as a function of temperature for selected values of $\hbar\omega$. The temperatures of the maxima, T(max), are indicated with an arrow for $\omega$ with the corresponding color. Panel e shows the same T(max) versus $\omega$. The solid curve is a linear fit extrapolating to $T(max)$=212 K for $\omega\rightarrow 0$. Spectra are shown in 10 meV intervals from 10 to 60 meV (150 meV) in panel c (d). Thick lines are used to highlight the data at selected energies. Data in the superconducting state in grey. The temperature range (370 K) of the lower panels is chosen such as to match the frequency range of the upper panels (200 meV) according to the scaling relation $2\pi k_B T = \hbar\omega$.

Turning now to $M_2(\omega,T)$ we observe that its frequency dependence exhibits an upward curvature for all temperatures. Also the temperature dependence has a $T^2$ component for the lowest frequencies. For frequencies higher than 50 meV the $T^2$ component is either absent or completely masked by the onset of superconductivity (grey segments of the temperature traces). Although $M_2(\omega,T)$ has no maximum as a function of temperature, the curves have an inflection point which shifts from roughly 200 K to 100 K when the frequency is raised from 10 to 50 meV. The saturation of $m^*(\omega,T)$ and the merging of the resistivity with a $T^2$ dependence indicate that the system enters a Fermi liquid like state at approximately 200 K. Plotting $M_2(\omega,T)$ as a function of $\omega^2$ (Fig. 4) we notice that for all temperatures above $T_c$, the initial rise is given by a linear slope as a function of $\omega^2$ (inset of Fig. 4). For a Fermi liquid $M(\omega,T)$ is in the relevant range of $\omega$ and $T$

$$M(\omega,T) \cong \left(\frac{1}{Z}-1\right)\hbar\omega + iC\left[(\hbar\omega)^2 + (p\pi k_B T)^2\right] \tag{2}$$

where $Z$ is the quasi-particle residue, $C$ is a constant with units of inverse energy, and $p=2$ for a local Fermi liquid or for three dimensions (see Appendix VIII). To check possible Fermi liquid characteristics of the data, we introduce a single parameter $\xi$ defined as $\xi^2=(\hbar\omega)^2+(p\pi k_B T)^2$, and we investigate $M_2$ as a function of $\xi$. As shown in Fig. 5 for three underdoped cuprate materials (Hg1201, ortho-II YBa$_2$Cu$_3$O$_{6.5}$ (Y123)[23], and Bi2201[24]) with hole concentration x≈0.1, the $M_2$ data of the normal state collapse in the low energy range on a single scaling curve for *p=1.5.* This value of p was obtained by searching for the best scaling collapse for *1≤p≤2* in steps of 0.1 (see Appendix IX*)*. Comparing the functional form of $M_2(\xi)$ for these three materials, we make the following observations: (i) Going from Hg1201 (left) to Bi2201 (right) in this plot, the residual ($\xi$ =0) value of $M_2(\xi)$ increases from 0 to about 80 meV. Indeed, it is believed that the relatively low values of $T_c$ in single layer Bi2201 have to do with strong scattering by disorder[25]. (ii) We also notice that in the case of Bi2201 some negative curvature shows up at the lowest energies, which is an indication that the Fermi liquid-characteristics are affected to some extent, and appear to be relatively fragile with respect to disorder. (iii) The implications of the loss of overlap above 100 meV in the Bi2201 data are not entirely clear. In principle there is no reason to expect overlap, since this is clearly beyond the range of "universal" Fermi liquid behavior. However, the single parameter scaling seems still to persist into this regime for the other two materials (Y123 and Hg1201), leading to the speculation that impurity scattering contributes to the disappearance of overlap above 100 meV for the Bi2201 sample.

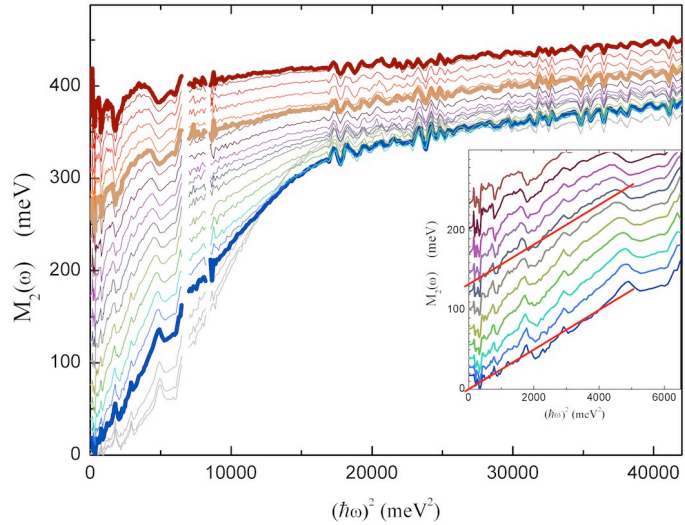

Figure 4 | Dynamical relaxation rate of underdoped Hg1201.
Imaginary part of the memory function of UD Hg1021 ($T_c$=67 K) for temperatures between 10 and 390 K in 20 K steps as a function of $\omega^2$. Thick lines are used to highlight the 70 K, 280 K and 390 K data. Inset: Zoom of the low-$\omega$ range showing a linear fit, temperatures are from 70 to 270 K in 20 K steps.

The most important observation borne out by these data is, that the frequency dependence of $M_1(\omega,T)$ and $M_2(\omega,T)$ follows by and large the behavior expected for a Fermi liquid : At low frequencies and temperatures $M_1(\omega,T)$ is indeed a linear function of $\omega$, and $M_2(\omega,T)$ scales with $(\hbar\omega)^2+(p\pi k_B T)^2$. We note that recent theories (*e.g.* Refs. 26-28) have emphasized the possible relevance of Fermi liquid concepts -or a hidden form of these[29]- to the metallic state of hole-doped cuprates. Our experimental observations provides a strong incentive for further theoretical work in this direction. We highlight two striking aspects of the data: (i) The slope $\partial M_1(\omega,T)/\partial\omega$ for $\omega\rightarrow0$ decreases significantly as a function of increasing temperature. (ii) *p<2.* We speculate that these issues are related to the progressive filling-in of the pseudo gap as a function of increasing temperature. Already in a two-fluid picture of a nodal Fermi liquid in parallel to an anti-nodal liquid non-universal features (for Fermi liquids) are introduced in the optical conductivity, since the properties at the Fermi surface change gradually from Fermi liquid at the nodes to strongly incoherent and pseudo-gapped at the hot spots near the anti-nodes. In fact, *p=2* has not been reported until now in any other material[30,31].

Theoretically it is expected that the $T^2$ and $\omega^2$ dependence of $M_2(\omega,T)$ is limited to $\hbar\omega$ and $p\pi k_B T$ lower than some energy scale $\xi_0$, which in the context of single parameter scaling behavior of a Fermi liquid is proportional to the effective Fermi energy. Strong electronic correlations strongly reduce this energy scale, as compared to the bare Fermi energy. For most materials the issue of the Fermi liquid like frequency dependence of $M_2(\omega,T)$ has remained largely unexplored. This is related to the difficulty that, in cases such as the heavy fermion materials where this type of coupling dominates, the range of Fermi liquid behavior is smaller than 10 meV, making it particularly difficult to obtain the required measurement accuracy in an infrared experiment. Clean underdoped cuprates present in this respect a favorable exception, since as can be seen from Figs. 4 and 5, the relevant energy scale $\xi_0$ is about 100 meV for a doping level around 10%. Above this energy $M_2(\omega,T)$ crosses over to a more linear trend both as a function of $\omega$ and $T$. This suggests that in cuprates the range of applicability of Fermi liquid behavior is limited by a different scattering mechanism that develops at high-$T$ and high-$\omega$, as the pseudogap gets filled.

The $\xi^2$ dependence of the relaxation rate can be understood as follows[1]: An electron at a distance $\xi$ above the Fermi energy can, as a result of electron-electron interactions, decay to a final state $\xi$-$\Omega$ by creating an electron-hole pair of energy $\Omega$. The density of states of electron-hole pairs is the spin (charge) susceptibility $\chi''(\Omega)$ where spin (charge) refers to electron-hole pairs carrying (no) net spin. $\chi''(\Omega)$ can be strongly renormalized, but the property that $\chi''(\Omega) \propto \Omega$ in the limit $\Omega \to 0$ is generic for Fermi liquids[8]. Integration of the susceptibility multiplied with the interaction vertex, $I^2\chi''(\Omega)$, over all possible decay channels from $0$ to $\xi$ leads us to conclude that indeed $M_2 \propto \xi^2$ as reported experimentally in the present manuscript. In this description the cross-over $\xi_0$ corresponds to the energy where $I^2\chi''(\Omega)$ is truncated, leading to a leveling off of $M_2$ for $\xi > \xi_0$. The strong temperature dependence of $M_1(\omega,T)$ is also a natural consequence of this description; it was shown in Ref. 32 that, in leading orders of temperature, $\chi''(\Omega)$ of a correlated Fermi liquid decreases as a function of temperature.

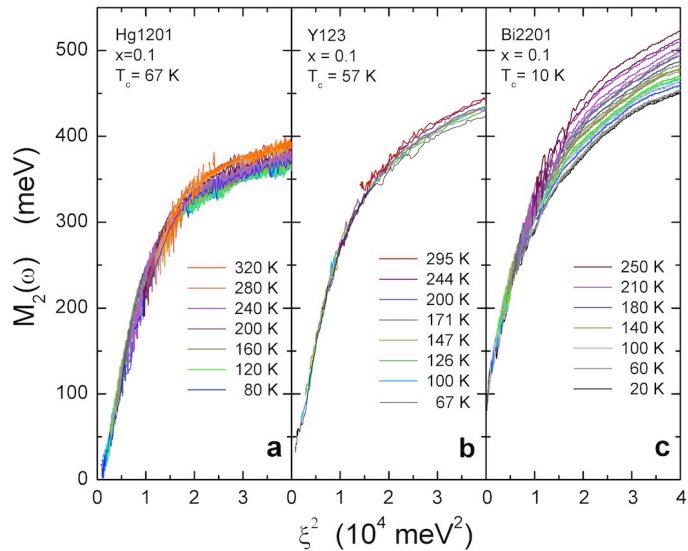

Figure 5 | Collapse of the frequency and temperature dependence of the relaxation rate of underdoped cuprate materials.
Normal state $M_2(\omega,T)$ as a function of $\xi^{2=}(\hbar\omega)^2+(p\pi k_B T)^2$ with $p=1.5$. a: Hg1201 ($x\cong 0.1$, $T_c$=67 K). b: Y123 ($x\cong 0.1$, $T_c$=57K), spectra by Hwang et al.[23] (digitized data of Fig. 6) represented here as a function of $\xi^2$, c: Bi2201 ($x\cong 0.1$, $T_c$=10K); data of van Heumen et al.[24] represented here as a function of $\xi^2$. The data displayed in panels a and c are in 10 K intervals with color coding indicated for temperatures in 40 K steps. In between these steps the color evolves gradually as a function of temperature. In panel b the color coding is given for all temperatures displayed.

In summary, we have revealed from optical spectroscopy measurements that the ungapped near-nodal excitations of underdoped cuprate superconductors obey Fermi liquid behavior when materials with reduced amount of disorder are considered. This observation, which is at variance with some established paradigms, provides new leads towards understanding of the metallic state and high-temperature superconductivity in these materials.

*Acknowledgements.* We are grateful to A. Chubukov, A. J. Leggett, T. Giamarchi, T. M. Rice, and T. Timusk for discussions and communications. This work was supported by SNSF through Grant No. 200020-140761 and the National Center of Competence in Research (NCCR). Materials with Novel Electronic Properties-MaNEP. The crystal growth and characterization work was supported by the US Department of Energy, office of Basic Energy Sciences. X.Z. acknowledges support by the National Natural Science Foundation, China. E.v.H. acknowledges support through the VENI program funded by the Netherlands Organisation for Scientific Research.

**Appendix I. Sample preparation**

Single crystals were grown using a flux method, characterized and heat treated to the desired doping level as described in Refs. 33 and 34. In Figs. A1-A6, 2-4 and 5a data are presented of a sample which has an onset critical temperature of 67 K and a transition width of 2 K. The crystal surface is oriented along the ab plane with a dimension of about 1.51 x 1.22 mm$^2$. Hg1201 samples are hygroscopic. Therefore the last stage of the preparation of the sample surface is done under a continuous flow of nitrogen, upon which the sample is transferred to a high vacuum chamber ($10^{-7}$ Torr) within a few minutes. Before each measurement the surface has been carefully checked for any evidence of oxidation.

**Appendix II. Comparison with DC resistivity**

Transport measurements have been performed using the 4-terminal method. Due to the irregular shape of the cleaved samples the absolute value of the DC resistivity be determined with about 20% accuracy. However, we obtained very high relative accuracy of the temperature dependance of the DC resistivity, as seen from identical temperature dependences of samples of the same composition and doping, regardless of having significantly different dimensions and shapes. An independent check of the DC resistivity was obtained from the *ω=0* limit of the experimental infrared optical conductivity (Figure 2 of main article). The DC-resistivity had to be scaled by a factor 0.66 in order to match the optical data, most likely due to aforementioned influence of the irregular shape of the crystals on the absolute value of the measured DC resistances. The excellent match of the two temperature dependencies demonstrates the high quality of both DC resistivity and optical conductivity data.

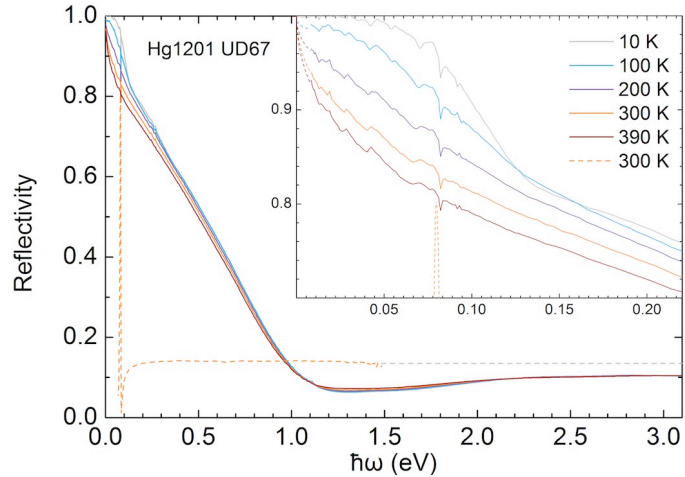

Figure A1 | Optical reflectivity of HgBa$_2$CuO$_{4+\delta}$ UD67 along the *a* (solid lines) and *c* (dashed line) axes at selected temperatures. A strong phonon mode is observed in both a and c axis at 82 meV. The inset shows the far infrared reflectivity where it is also possible to see the remaining of a suppressed *c*-axis phonon at 70 meV.

# Appendix III. Optical spectroscopy

The infrared reflectivity of the ab plane was measured at near-normal incidence using a Fourier-transform spectrometer (8 meV – 1.24 eV). In addition, ellipsometric measurements were performed in the near-IR to near-UV using a Woolam VASE elipsometer (0.8 eV – 3.7 eV). The procedure of Ref. 35 was used to suppress possible spurious c-axis components in the reflectance curves. The absolute value of the reflectivity was obtained by an in situ gold evaporation on the sample. The sample was mounted in a high vacuum home-made cryostat designed for high stability during the thermal cycles, a prerequisite for absolute temperature dependence of the reflectivity. The operating pressure was $10^{-7}$ mbar. Temperature cool down sweeps have been performed between 395 and 10K at a speed of 0.9 K/min leading to about one reflectivity spectrum per Kelvin. In order to increase the signal to noise ratio, the data has been binned in 10K temperature intervals. The long time-scale drift of the mid-infrared detectors has been calibrated using a flipping mirror placed outside the cryostat in front of the sample. The absolute reflectivity calibrated for spectrometer throughput and drift is obtained from the relation

$$R(\omega) = \frac{I_{sample}(\omega)}{I_{reference}(\omega)} \frac{I_{reference-mirror}(\omega)}{I_{sample-mirror}(\omega)} \tag{A1}$$

The reflectivity spectra for selected temperatures are displayed in Fig. A1. The consistency of this procedure is confirmed by the good correlation with the resistivity measurement shown in Fig. 2.

In addition, the c-axis reflectivity was measured at room temperature on a polished edge of the sample using an infrared microscope attached to a conventional Fourier transform spectrometer. The frequency range of that measurement goes from 68 meV to 1.5 eV.

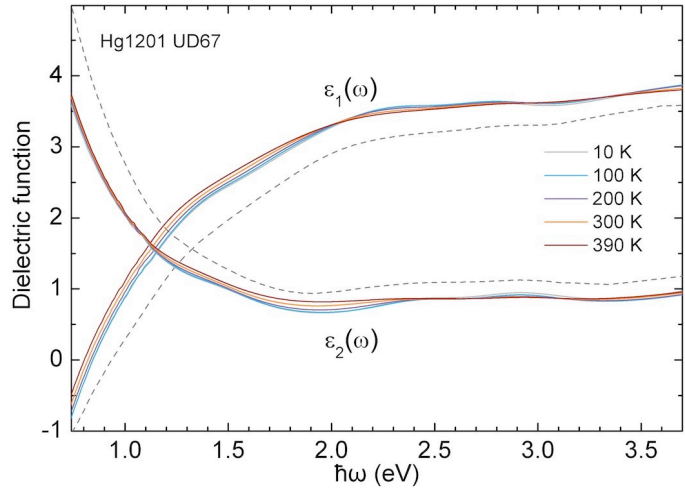

Figure A2 | Real and imaginary parts of the pseudodielectric function (dashed curves) obtained from ellipsometry at room temperature, and *ab*-plane dielectric function at selected temperatures corrected using the method explained in the text.

For frequencies above 0.8 eV we used spectroscopic ellipsometry with an angle of incidence set at 61 degrees relative to the surface normal, providing after inversion of the Fresnel equations the so-called pseudo dielectric function

$$\varepsilon_{ps}(\omega) = \sin^2\theta + \sin^4\theta \left( \frac{\sqrt{\varepsilon_{ab} - \sin^2\theta}\sqrt{\varepsilon_{ab}} - \sqrt{1 - \varepsilon_c^{-1}\sin^2\theta}}{\sqrt{\varepsilon_{ab} - \sin^2\theta}\sqrt{1 - \varepsilon_c^{-1}\sin^2\theta} - \sqrt{\varepsilon_{ab}}\cos^2\theta} \right)^2 \tag{A2}$$

where $\varepsilon_{ab}$ and $\varepsilon_c$ are the dielectric tensor elements in the ab-plane and along the c-axis respectively. The red dotted curves in Fig. A2 represent the pseudo dielectric function measured at 300 K. In the limit of zero anisotropy ($\varepsilon_{ab}=\varepsilon_c$) the pseudo-dielectric function becomes the dielectric function ($\varepsilon_{ps}=\varepsilon_{ab}$). We corrected for the c-axis dielectric function by fitting the pseudo-dielectric function, the ab-plane reflectance and the c axis reflectivities to a Drude-Lorentz model. The resulting ab-plane dielectric functions are shown in Fig. A2 for a few selected temperatures. The reflectance and phase spectra were calculated in this data range using the

Fresnel expression for normal incidence reflectivity. A perfect match was obtained in the range of overlap (0.8-1.2 eV) of ellipsometry and direct normal incidence reflectance data. The complex dielectric constant was obtained by standard Kramers-Kronig transformation of the reflectance data, using the reflectance *and* phase data between 0.8 and 3.7 eV to fix the high frequency extrapolation. This procedure anchors the phase output of the Kramers-Kronig transformation in the entire frequency range to the experimental data between 0.8 and 3.7 eV.

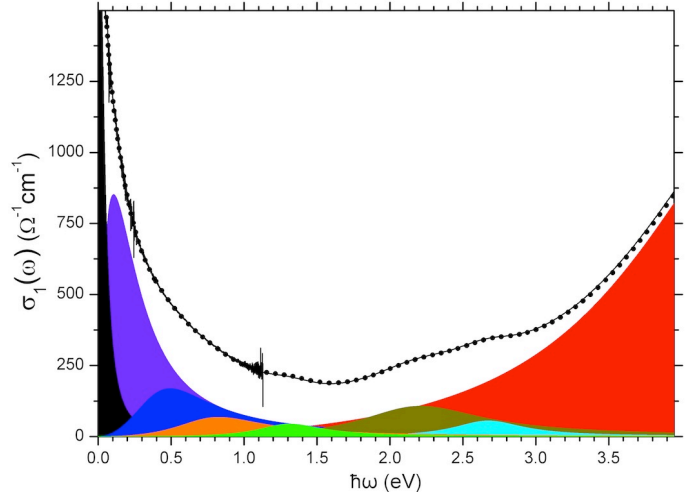

Figure A3 | The Drude-Lorentz fit to the real part of the optical conductivity, $\sigma_1(\omega)$, at 300 K. The 4 Lorentzians above 1.2 eV are interpreted as bound charge contributions to $\sigma_1(\omega)$. The dotted grey curve is the sum of all oscillators which matches perfectly the $\sigma_1(\omega)$ calculated from Kramers-Kronig transformation (black thin curve).

**Appendix IV. Drude Lorentz analysis**

In order to characterize the oscillator strengths and frequencies corresponding to the interband transitions, we have fitted to the experimental conductivities a linear superposition of Drude and Lorentz oscillators

$$\varepsilon(\omega) = 1 + S_h + \sum_{j=0}^{N} \frac{\omega_{pj}^2}{\omega_j^2 - \omega(\omega + i\gamma_j)} \tag{A3}$$

where $S_h$ summarizes the dielectric polarizability originating in oscillators at frequencies higher than the $j=7$ mode at 5.2 eV. The dielectric function is understood to represent the superposition of the conduction electron (or hole) optical conductivity and the "bound charge" response

$$\varepsilon(\omega) = \frac{4\pi i}{\omega}\sigma_f(\omega) + \varepsilon_b(\omega) \tag{A4}$$

The optical conductivity is shown in Fig. A3, together with the Drude-Lorentz fit. The corresponding parameters are summarized in Table T1.

|  | j=0 | j=1 | j=2 | j=3 | $\hbar\omega_p$ | j=4 | j=5 | j=6 | j=7 | $S_h$ |
|---|---|---|---|---|---|---|---|---|---|---|
| $\hbar\omega_{0,j}$ | 0.000 | 0.116 | 0.542 | 0.908 |  | 1.443 | 2.430 | 2.939 | 5.204 |  |
| $\hbar\omega_{p,j}$ | 0.857 | 1.461 | 0.969 | 0.638 | 2.053 | 0.444 | 0.947 | 0.510 | 4.390 | 2.2 |
| $\hbar\gamma_j$ | 0.046 | 0.337 | 0.747 | 0.805 |  | 0.598 | 1.126 | 0.636 | 2.723 |  |

TABLE T1. Parameters of the Drude-Lorentz oscillators displayed in Fig. A3. $S_h$ is dimensionless. The parameter values corresponding to $\hbar\omega_p$, $\hbar\omega_0$, and $\hbar\gamma$ are in units of eV.

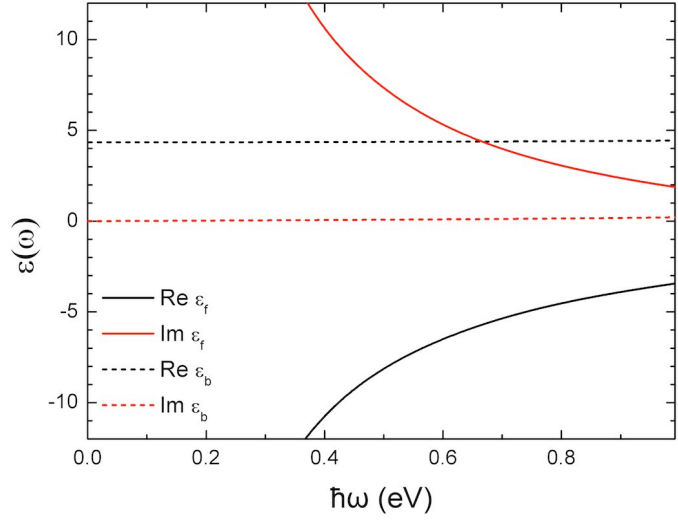

Figure A4 | Total (solid curves) and bound charge dielectric function (last four oscillators of Table T1) at room temperature. To a very good approximation $\varepsilon_b=4.3\pm0.1$ with negligible imaginary part in the entire frequency range shown.

**Appendix V. Conduction band and bound charge contributions**

Here we will interpret all oscillators below 1.2 eV as an intrinsic part of the conduction band response. The interpretation of the weak green peak at 1.4 eV is uncertain. Since the spectral weight is very small, it makes next to no difference for the discussion in the present paper whether or not we assume it to be part of the conduction band response. A reasonable approximation for the total (coherent + incoherent) conduction band spectral weight is therefore

$$\omega_p^2 = \sum_{\hbar\omega_j<1.2eV} \omega_{pj}^2 \qquad (A5)$$

The resulting value for $\hbar\omega_p$ is indicated in table T1. An alternative scheme for determining the conduction band spectral weight consists of fitting the dielectric function in the full frequency range to a sum of bound charge oscillators above 1.2 eV and to use Allen's formula for the conduction band conductivity of electrons coupled to bosons[18,24,36], where in the latter the amplitudes of the blocks of a histogram representation of the electron-boson coupling function are adjusted to obtain the best fit.

The rise in optical conductivity above 1.8 eV is due to the onset of O2p→Cu3d charge transfer transitions. The dielectric response described by the oscillators above 1.2 eV is therefore interpreted as "bound charge" response. The bound charge component of the dielectric function is

$$\varepsilon_b(\omega) = 1 + S_h + \sum_{\hbar\omega_j>1.2eV} \frac{\omega_{pj}^2}{\omega_j^2 - \omega(\omega+i\gamma_j)} \cong 1 + S_h + \sum_{\hbar\omega_j>1.2eV} \frac{\omega_{pj}^2}{\omega_j^2} \qquad (A6)$$

The right hand side of the expression is a valid approximation for the present set of data. This is demonstrated in Fig. A4 showing Re $\varepsilon_b(\omega)$ and Im $\varepsilon_b(\omega)$ using the parameters in table T1, together with the total dielectric function. We see, that $\varepsilon_b(\omega)=4.3\pm0.1$ everywhere in the frequency range shown.

**Appendix VI. Sheet conductance and spectral weight**

We consider a cuprate with $N_L$ conducting $CuO_2$ sheets per unit cell, having c-axis lattice parameter $c$. This corresponds to an average spacing $d_c=c/N_L$ between the $CuO_2$ sheets. The relation between sheet conductance and bulk conductivity is

$$G(\omega) = d_c \sigma(\omega) \tag{A7}$$

The general expression for spectral weight of the sheet conductance along the $x_j$ axis is[18]

$$K = \frac{d_c}{V} \sum_{k,\sigma} n_{k,\sigma} \frac{\partial^2 \varepsilon_{k,\sigma}}{\partial k_j^2} \tag{A8}$$

where $n_{k\sigma}$ is the average occupation of the state with momentum $k$ and spin $\sigma$. $V$ is the sample volume. For a parabolic dispersion relation ($\varepsilon_k = \hbar^2 k^2/2m$) the Fermi energy is $\pi K$. Ab initio LDA band-calculations for the cuprates can be accurately represented by the tight-binding expression[37]

$$\varepsilon_k = -2t\left[\cos(k_x a) + \cos(k_y a)\right] + 4t'\cos(k_x a)\cos(k_y a) - 2t''\cos(2k_x a)\cos(2k_y a) \tag{A9}$$

We calculated $K$ numerically as a function of hole density, inserting Eq. A9 in Eq. A8, and adopting the parameter set appropriate to the case of Hg1201[37]: $t=0.45$ eV, $t'/t=0.35$, $t''/t'=0.5$. The parametric plot of $K$ versus hole density is presented in Fig. 1 (black solid curve).

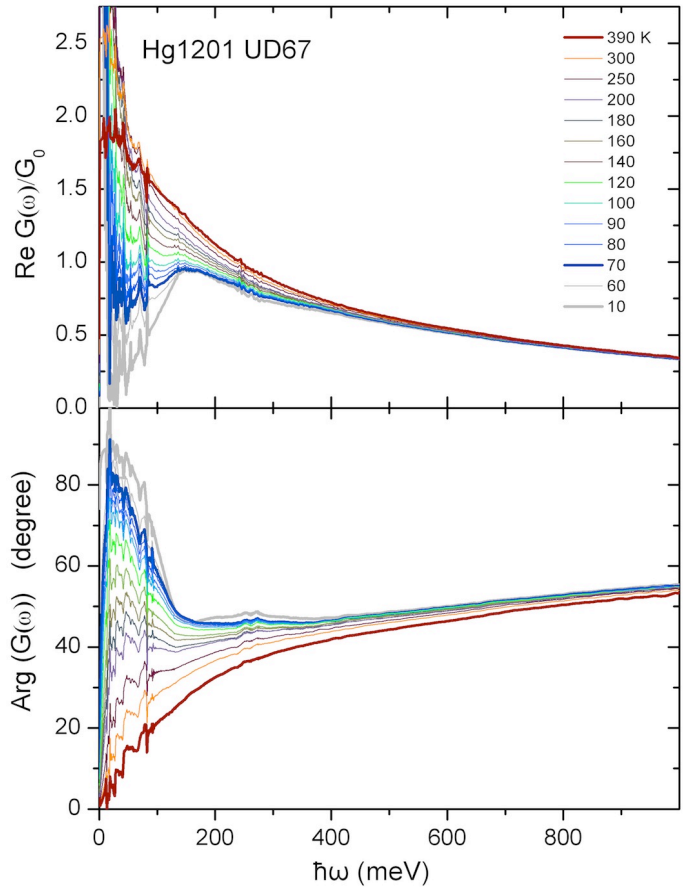

Figure A5 | Optical sheet conductance and the corresponding phase of UD Hg1201. Temperatures from 10 to 390 K. Data below $T_c$ are shown in grey.

**Appendix VII. Generalized Drude analysis**

The real part of the optical conductivity is shown in the upper panel of Fig A5 for different temperatures. There is a clear signature of a pseudogap above $T_c$, with a maximum at approximately 0.15 eV. The phase of conduction band optical conductivity, shown in the lower panel of Fig A5, has a gradual rise of phase as a function of frequency, in agreement with the behavior of underdoped Bi2212 pointed out in Ref. 2, and distinctly different from $\sigma(\omega) = C(i\omega)^{-2/3}$ observed in optimally doped cuprates.

A generally used expression for the optical conductivity of a liquid of interacting electrons is the so-called extended Drude formula, expressed in the context of this article as a sheet conductance in units of $G_0=2e^2/h$

$$G(\omega) = \frac{i\pi K}{\hbar\omega + M(\omega)} G_0 \qquad (A10)$$

where $M(\omega,T)$ is defined as a "memory function"[10] encoding the departure from standard Drude behaviour caused for instance by electron-electron interactions or electron-phonon coupling. For a Fermi liquid, the imaginary part of $M(\omega,T)$ has a simple interpretation as the relaxation rate of the unrenormalized electrons, Im $M(\omega,T)=\hbar/\tau(\omega,T)$, and Re $M(\omega,T)/\hbar\omega+1=m^*(\omega,T)/m$ is the frequency dependent mass renormalization.

The spectral weight factor $K$ is related to the plasma frequency through the relation

$$\hbar^2\omega_p^2 = \frac{4\pi e^2}{d_c} K \qquad (A11)$$

With $\hbar\omega_p=2.053$ eV (table T1) and the c-axis parameter $d_c=0.952$ nm we obtain $K=222$ meV. The memory functions in Figs. 3, 4 and 5 have been calculated from the optical conductivity using Eq. A10. The sharp features at 91, 82 meV at 45 meV and 35 meV of the Hg1201 data are due to dipole active optical phonons. Although in principle these could be subtracted from the optical conductivity before calculating $M(\omega,T)$, this kind of subtraction procedure contains some ambiguities due to Fano-like phonon asymmetries. Since after subtraction these ambiguities are imported in the resulting memory function, we have refrained from subtracting the phonons. Instead, in order to facilitate comparison of different temperatures, in Figs. 4 and 5 narrow bands corresponding to the prominent sharp features at 82 and 91 meV of the UD Hg1201 data have been left open, as can be seen in the top panel of Fig. 4 of the main text.

**Appendix VIII. Memory function of a local Fermi liquid**

For a system whose carriers can be well described by a local (i.e. momentum independent) scattering rate, the optical conductivity can be expressed entirely in terms of the complex electronic self-energy $\Sigma(\varepsilon)\equiv\Sigma_1(\varepsilon)+i\Sigma_2(\varepsilon)$. The scattering rate is $-\Sigma_2$, and $\Sigma_1$ is linked to $\Sigma_2$ by a Kramers-Kronig transformation. This happens because, for a local self-energy, the vertex corrections vanish in the Kubo formula for the conductivity $\sigma(\omega)$. As a result,

$$\sigma(\omega,T) = \int d\varepsilon \frac{i\Phi(0)}{\omega} \frac{f(\varepsilon)-f(\varepsilon+\hbar\omega)}{\hbar\omega + \Sigma^*(\varepsilon) - \Sigma(\varepsilon+\hbar\omega)} \qquad (A12)$$

Here $f(\varepsilon)=[exp(\varepsilon/k_BT)+1]^{-1}$, and $\Phi(0)$ is proportional to the Fermi-surface average of the square of the bare electronic group velocity. In a Fermi liquid, the quasi-particles are well-defined at low temperature close to the Fermi surface, because in this limit the scattering rate vanishes with a characteristic energy and temperature dependence proportional to $\varepsilon^2+(\pi k_BT)^2$. The quasi-particle spectral weight $Z$ is smaller than unity, as reflected by the fact that $\Sigma_1(\varepsilon)$ has a finite negative slope
$1-1/Z$ at $\varepsilon=0$. This leads to the following low-energy model for the self-energy of a local Fermi liquid:

$$\Sigma(\varepsilon) = \left(1 - \frac{1}{Z}\right)\varepsilon - \frac{3}{2}iC\left[\varepsilon^2 + \pi^2(k_BT)^2\right] \qquad (A13)$$

The added factor 3/2 anticipates the formula for the memory function. In systems where all properties are set by a single characteristic energy $D$, like in isotropic doped Mott insulators where $D$ is typically the half bandwidth, the coefficient $C$ would be proportional to $1/(Z^2D)$. Indeed, in such cases the scattering rate $-\Sigma_2(0)$ scales with renormalized electronic energies like $D(k_BT/ZD)^2$. The quasi-particle life-time controls the Drude response at very low frequencies. On the Fermi surface, it can be computed from the self-energy as $\tau_{qp}=\hbar/(2Z|\Sigma_2(0)|)$. We are interested in the regime of frequencies much larger than $1/\tau_{qp}$: introducing Eq. (A13) into Eq. (A12), and expanding for large $\omega\tau_{qp}$, we get

$$\sigma(\omega,T) = \Phi(0)Z\tau_{qp}\left\{\frac{i}{\omega\tau_{qp}} + \frac{4}{3(\omega\tau_{qp})^2}\left[1+\left(\frac{\hbar\omega}{2\pi k_BT}\right)^2\right]\right\} + O\left[1/(\omega\tau_{qp})^3\right] \qquad (A14)$$

The expression of the conductivity in terms of the memory function is $\sigma(\omega)=i\hbar\Phi(0)/[\hbar\omega+M(\omega)]$. Comparing this with Eq. (A13), we obtain Eq. (2) of the main text in the regime $\omega\tau_{qp}>>1$. Note that here, for consistency with Eq. (A11), we have defined $M(\omega)$ using the full spectral weight, which is proportional to $\Phi(0)$. If the experimentally determined spectral weight K underestimates the full spectral weight by a factor $\alpha$, the measured value of the residue -from the slope of $M_1(\omega)$- will overestimate Z by a factor $\alpha$.

**Appendix IX. Scaling collapse**

We tested the frequency and temperature dependence of the imaginary part of the memory function for scaling collapse of the form of Eq. 2 of the main text by plotting of the data in the normal state as a function of $\xi^2=(\hbar\omega)^2+(p\pi k_B T)^2$, and searching for the value of $p$ that provides the best overlap of the low-$\xi$ data for the spectra taken at different temperatures. As we see in Fig. A6, neither for $p=1$, nor for $p=2$ the data collapse on a single curve. The optimal value where the data does collapse, $p=1.5$, is used in Fig. 5. Note that de data above 320 K correspond to the plateau at high frequencies, which is clearly outside the range of validity of single-parameter scaling. For this reason the data shown in Fig 5 was restricted to the temperatures below 320 K.

In Ref. 2 a different scaling relation -pertaining to the real part of the optical conductivity- was reported

$$\frac{\hbar}{k_B T \sigma_1(\omega,T)} = \frac{4\pi}{\omega_p^2}\left(1+A^2\left(\frac{\hbar\omega}{k_B T}\right)^2\right) \quad (A15)$$

where $A$ is a number of order 1. This is the classical relaxation dynamics expected near a quantum critical point[39]. One may wonder whether this behavior can occur together with single parameter scaling of the form $M_2(\omega,T)=M_2(\xi)$, i.e. similar to Eq. 2 of the main text but not necessarily corresponding to a $\xi^2$-dependence of $M_2$. The aforementioned scaling relation of $T\sigma_1(\omega,T)$ can be obtained if $M_2(\omega,T)=iA^{-1}k_B T$ for $\omega/T\rightarrow 0$. Since in this limit $\xi=k_B T$, consequently $M_2$ has a linear dependence on $\xi$, at least for not too large $\xi$. However, lacking a theoretical justification for scaling of the form $M_2(\omega,T)=M_2(\xi)$ for systems which are not a Fermi liquid, we restrict the present discussion to the underdoped cuprates in the pseudogap phase.

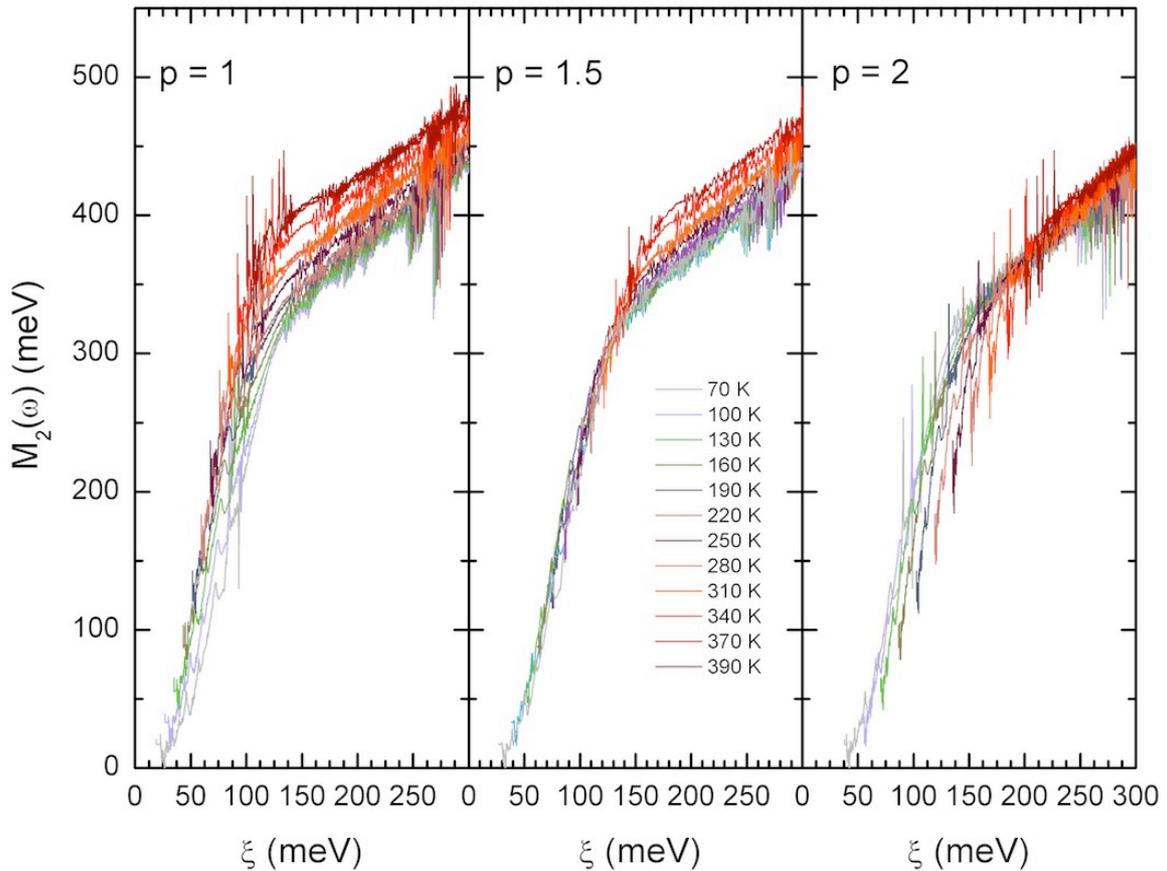

Figure A6 | Imaginary part of the memory function as a function of $\xi$, where $\xi^2=(\hbar\omega)^2+(p\pi k_B T)^2$ with $p=1$ (a), $p=1.5$ (b), and $p=2$ (c) for underdoped Hg1201 ($T_c=67$ K). The data corresponds to the normal state.